\documentclass[twocolumn,showpacs,amsmath,amssymb,aps,prb,superscriptaddress,tbtags]{revtex4-2}

\usepackage[]{graphicx}
\usepackage[]{verbatim}
\usepackage[]{color}
\usepackage{overpic}
\usepackage{rotating}
\usepackage{mathtools}
\usepackage{appendix}
\usepackage{ulem}
\usepackage[margin=0.6in]{geometry}
\usepackage{tabularx}
\usepackage{float}
\usepackage{multirow}

\usepackage[]{hyperref}

 \hypersetup{
 colorlinks = true,
 linkcolor = blue,
 urlcolor = blue,
 citecolor = blue}

\usepackage{array}
\newcolumntype{L}[1]{>{\raggedright\let\newline\\\arraybackslash\hspace{0pt}}m{#1}}
\newcolumntype{C}[1]{>{\centering\let\newline\\\arraybackslash\hspace{0pt}}m{#1}}
\newcolumntype{R}[1]{>{\raggedleft\let\newline\\\arraybackslash\hspace{0pt}}m{#1}}
\newcolumntype{N}{@{}m{0pt}@{}}

\makeatletter
\newsavebox{\@brx}

\newcommand{\llangle}[1][]{\savebox{\@brx}{\(\m@th{#1\langle}\)}%
  \mathopen{\copy\@brx\mkern2mu\kern-0.8\wd\@brx\usebox{\@brx}}}
\newcommand{\rrangle}[1][]{\savebox{\@brx}{\(\m@th{#1\rangle}\)}%
  \mathclose{\copy\@brx\mkern2mu\kern-0.8\wd\@brx\usebox{\@brx}}}

  \newcommand{\lllangle}[1][]{\savebox{\@brx}{\(\m@th{#1\langle}\)}%
  \mathopen{\copy\@brx\copy\@brx\mkern4mu\kern-0.7\wd\@brx\usebox{\@brx}}}
\newcommand{\rrrangle}[1][]{\savebox{\@brx}{\(\m@th{#1\rangle}\)}%
  \mathclose{\copy\@brx\copy\@brx\mkern4mu\kern-0.7\wd\@brx\usebox{\@brx}}}

\graphicspath{{./}{./}}

\begin{document}
\title{Evolution of Interorbital Superconductor to Intraorbital Spin-Density Wave in Layered Ruthenates}
\author{Austin W.~Lindquist}
\affiliation{Department of Physics and Center for Quantum Materials, University of Toronto, 60 St.~George St., Toronto, Ontario, M5S 1A7, Canada}
\author{Jonathan Clepkens}
\affiliation{Department of Physics and Center for Quantum Materials, University of Toronto, 60 St.~George St., Toronto, Ontario, M5S 1A7, Canada}
\author{Hae-Young Kee}
\email{hykee@physics.utoronto.ca}
\affiliation{Department of Physics and Center for Quantum Materials, University of Toronto, 60 St.~George St., Toronto, Ontario, M5S 1A7, Canada}
\affiliation{Canadian Institute for Advanced Research, Toronto, Ontario, M5G 1Z8, Canada}
\begin{abstract}
The ruthenate family of layered perovskites has been a topic of intense interest, 
with much work dedicated to the superconducting state of Sr$_2$RuO$_4$.
Another longstanding puzzle is the lack of superconductivity in its 
sister compound, Sr$_3$Ru$_2$O$_7$, which constrains the possible
mechanisms of Sr$_2$RuO$_4$.
Here we address a microscopic mechanism that 
unifies the orders in these materials.
Beginning from a model of Sr$_2$RuO$_4$ featuring interorbital spin-triplet pairing
via Hund's and spin-orbit couplings,
we find that bilayer coupling alone enhances, 
while staggered rotations destroy interorbital superconductivity.
A magnetic field then shifts van Hove singularities,
allowing intraorbital spin-density wave order to form in Sr$_3$Ru$_2$O$_7$.
Our theory predicts that 
Sr$_3$Ru$_2$O$_7$ without staggered rotations exhibits 
interorbital superconductivity with a possibly higher transition temperature.
\end{abstract}
\maketitle

\section{Introduction}
Multiorbital systems in which the low-energy behavior is affected by  
various orbitals provide an ideal setup for the formation of rich 
electronic phases. 
One notable example is the family of layered perovskites, Sr$_{n+1}$Ru$_n$O$_{3n+1}$.
Much attention has been paid to the single layer material, Sr$_2$RuO$_4$ (214), which 
displays superconductivity (SC) at low temperatures \cite{Maeno1994Nature,Mackenzie2003RMP,Kallin2012RPP,Mackenzie2017NPJ}.
However, despite intense studies, the nature of the superconducting state 
still remains a topic of debate \cite{Mackenzie2017NPJ,Kivelson2020npj}.
Interest in the bilayer material, Sr$_3$Ru$_2$O$_7$ (327), grew after 
significant anisotropy in the resistivity, signaling an electronic
nematic phase, was reported 
\cite{Perry2001PRL,Grigera2001science,Perry2004PRL,Grigera2004science,Kitagawa2005PRL,Borzi2007science,Mackenzie2012PC}.
More recently, a resistivity linear in temperature \cite{Bruin2013science},
as well as spin-density wave (SDW) order were found to occur at low temperatures
at field values coinciding with the 
presumed nematic phase, bounded by metamagnetic transitions \cite{Berridge2009PRL, Berridge2010PRB, Lester2015nmat, Lester2021nc}.

A longstanding puzzles in this family is the lack of SC in 327,
constraining the
possible superconducting mechanisms of 214.
The similarity of these materials means that a model of 214
should be able to explain the lack of SC in 327
and the sensitivity to the differences between the two.
The two notable differences 
introduced by the second layer in 327 
are the bilayer coupling and the staggered rotations of the 
oxygen octahedra.  
The observation of SDW order in the presence of a magnetic field in 327
suggests a delicate competition between 
SC and SDW in
these two sister compounds, 
calling for a unified picture.  

\begin{figure}[t]
\includegraphics{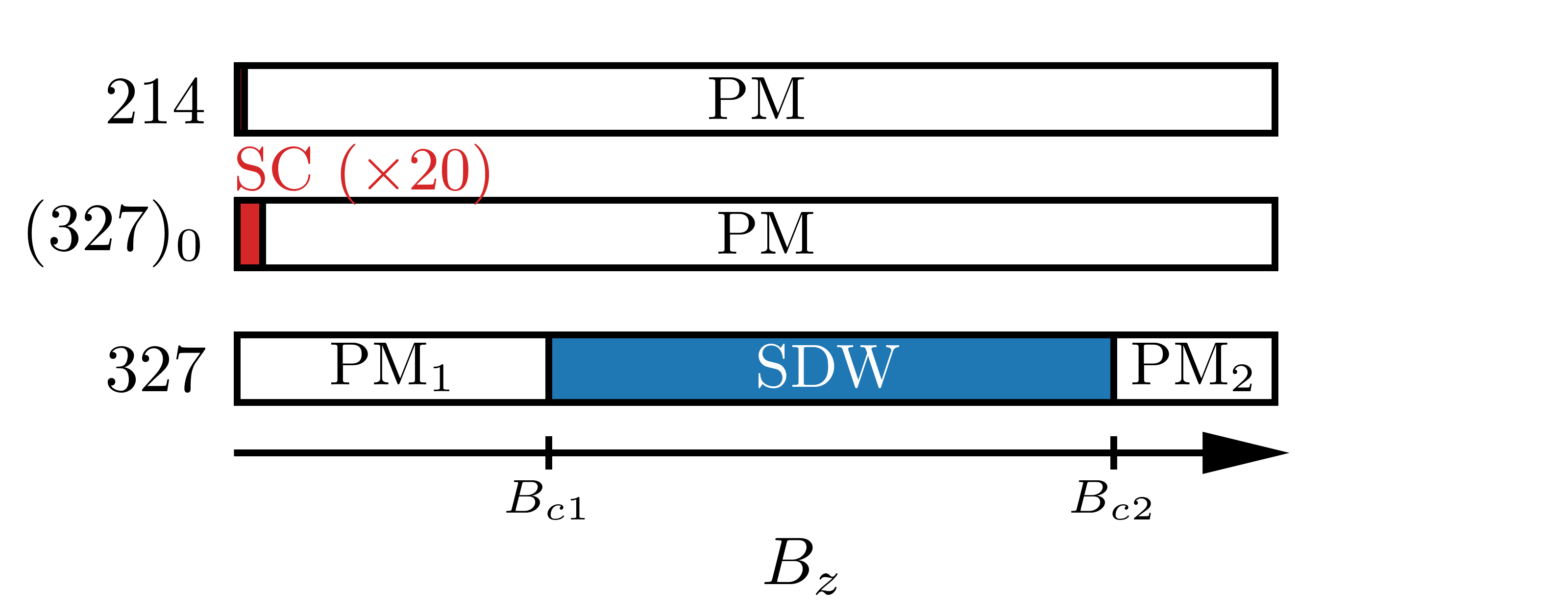}
\caption{
Phase diagram of the single-layer, 214, ideal bilayer (no staggered rotations), $(327)_0$, and real bilayer, 327.
214 features superconductivity at small fields,
expanded in $(327)_0$.
Both regions are expanded by a factor of 20 to more clearly show their relative sizes.
In 
327, SC is destroyed by rotations of the octahedra, 
and SDW order exists at finite field values between $B_{c1}$ and 
$B_{c2}$. 
The values of $B_{c1}$ and $B_{c2}$ and other relevant parameters are given in the main text.
The two distinct paramagnetic regions surrounding the SDW order, labeled PM$_1$ and PM$_2$,
represent the low- and high-moment regions, respectively.
}\label{pd}
\end{figure}

Here, we address the microscopic mechanism of the evolution from SC
in 214 to  
SDW  
in 327.
We show that the absence of SC in 327 supports interorbital 
pairing in 214 which is sensitive to 
orbital mixing via spin-orbit coupling (SOC).
Using the same parameter set for the Kanamori Hamiltonian, 
we find interorbital spin-triplet SC is destroyed by the staggered 
rotations of the octahedra, and intraorbital SDW order emerges under
a magnetic field in 327.
Further, we predict that ideal bilayer 327 without staggered rotations of the
octahedra, denoted $(327)_0$, displays interorbital SC with a transition temperature 
possibly higher than 214, as shown in Fig.~\ref{pd}, the phase diagram of these
systems in a field.

\section{Microscopic theory}
Sr$_2$RuO$_4$ was once thought to be a leading candidate for $p+ip$ spin-triplet SC 
\cite{Rice1995JPCM,Ishida1998nature}. 
However, recent NMR data showing a drop in the Knight shift below the superconducting transition temperature seems to rule out
odd-parity spin-triplet pairing \cite{Pustogow2019Nature,Chronister2021pnas}.
While $p$-wave states may not explain the Knight shift reduction,
spin-triplet proposals 
remain possible with even-parity interorbital spin-triplet states
\cite{Puetter2012EPL, Hoshino2015PRL, Hoshino2016PRB, Gingras2019PRL, Suh2020PRR, Lindquist2020PRR, Clepkens2021PRR, Clepkens2021PRB}.
This pairing in the orbital basis is spin-triplet,
but appears as pseudospin-singlet
in the band basis due to 
SOC.
The interorbital nature of the pairing means that significant 
band degeneracy near the Fermi level is required 
\cite{Klejnberg1999iop,Dai2008PRL}.
Alternatively, when SOC 
and band separation are comparable in energy, 
SOC stabilizes interorbital pairing via mixing of 
orbitals and spin 
\cite{Puetter2012EPL,Veenstra2014PRL,Ramires2016PRB,Vafek2017PRL,Cheung2019PRB,Ramires2019PRB,Suh2020PRR,Clepkens2021PRR,Clepkens2021PRB}.

We adopt the Kanamori Hamiltonian, well known for multiorbital systems, 
to investigate competition between SC and SDW order in 214,
$(327)_0$, and 327.
In multiorbital systems, 
the distinction between intraorbital,
interorbital-singlet, and interorbital-triplet pairings must all be considered.
Among all possible order parameters, inter- and intraorbital SDW, and interorbital
SC occur via attractive interactions, however, interorbital SDW is small compared to
intraorbital SDW and is therefore not shown.
Intraorbital SC and charge-density wave channels feature repulsive interactions,
but intraorbital SC is induced from interorbital SC via SOC, though is small in
comparison.

Taking into account 
interorbital SC and intraorbital SDW, the
effective Hamiltonian is found as,
\begin{equation}\label{Hint}
\begin{aligned}
\frac{H_\text{eff}}{N} =&~2(U'-J_H)\sum_{a\ne b} \hat{\vec{\Delta}}_{a/b}^\dagger \cdot \hat{\vec{\Delta}}_{a/b} \\
&-2U\sum_a \hat{\vec{m}}_\textbf{q}^a\cdot\hat{\vec{m}}_{-\textbf{q}}^a
-J_H\sum_{a\ne b}\hat{\vec{m}}_\textbf{q}^a\cdot\hat{\vec{m}}_{-\textbf{q}}^b.
\end{aligned}
\end{equation}
The orbital-singlet, spin-triplet, superconducting order parameter, $\hat{\Delta}_{a/b}^{l\dagger}$, is written as,
\begin{equation}\label{Delta}
\hat{\Delta}_{a/b}^{l\dagger}=\frac{1}{4N}\sum_{\bf k}[i\hat{\sigma}^y \hat{\sigma}^l]_{\sigma \sigma'}(c_{{\bf k},\sigma}^{a\dagger} c_{-{\bf k},\sigma'}^{b\dagger}-c_{{\bf k},\sigma}^{b\dagger} c_{-{\bf k},\sigma'}^{a\dagger}),
\end{equation}
with $l=x,y,z$ and $a\ne b$ represents the sum over the unique pairs of orbital indices in $(yz,xz,xy)$.
The intraorbital SDW order parameter, $\hat{\vec{m}}_\textbf{q}^a$, is given by,
\begin{equation}\label{OP}
\hat{\vec{m}}_\textbf{q}^a=\frac{1}{2N}\sum_k c_{\textbf{k}+\textbf{q}/2,\sigma}^{a\dagger}[\vec{\sigma}]_{\sigma,\sigma'}c_{\textbf{k}-\textbf{q}/2,\sigma'}^a,
\end{equation}
where the electron operator $c_{\textbf{k},\sigma}^{a\dagger}$ creates an electron in
orbital $a$ with spin $\sigma$.

To investigate the competition between SC and SDW order in these three 
systems, we must determine their tight-binding (TB) Hamiltonians.
TB parameters are obtained from density functional theory (DFT) calculations,
and listed in Tables AI and AII of appendix A.
The SC pairing itself is insensitive to the choice of TB parameters, however, the 
symmetry of the pairing state depends on the details as shown in Ref.~\onlinecite{Clepkens2021PRB}. 
While the pairing in the orbital basis has no momentum dependence, higher-angular momentum pairing can
be revealed in the band basis due to momentum-dependent SOC or
a combination of dispersion terms and SOC 
\cite{Clepkens2021PRR, Clepkens2021PRB}. 
The $A_{1g}$ pairing is favored by the atomic SOC considered here, however, 
all irreducible representations of the $D_{4h}$ point group are possible
from interorbital SC \cite{Huang2019PRB, Kaba2019PRB}, with others favored by momentum-dependent SOCs
\cite{Suh2020PRR, Clepkens2021PRR, Clepkens2021PRB}. 
This allows for gap nodes as well as a complex multi-component order parameter
as suggested by experiments \cite{Benhabib2020, Ghosh2020},
beyond the current study. 
Next, we explore the effects of orbital-dependent bilayer
coupling on SC. We also examine hopping terms from staggered rotations, 
and show that this is responsible for the lack of SC in 327.

\section{Evolution of SC}
Self-consistent mean-field approximation (MFA) calculations are performed 
for the TB models of 214, (327)$_0$, and 327.
(327)$_0$ uses the TB parameters from the single layer with interlayer hoppings added, 
referred to collectively as the bilayer coupling.
In addition to bilayer coupling, 
intralayer hopping terms may be modified or added
in 327
due to the staggered rotations of octahedra 
not present in  the single layer.

The phase diagram for each of the three systems
is shown in Fig.~\ref{pd} as functions of the effective
MFA interactions, $V_\text{SDW} \equiv U$ and $V_\text{SC}\equiv J_H-U'$.
Comparing with 214, it is clear that the SC region is extended in (327)$_0$, 
while the SDW region is extended in 327.
Since the precise values of $U$, $U'$, and $J_H$ are unknown, we set $U'=U-2J_H$, 
and choose
a set of $(U,J_H)$ which gives SC in 214.  These same values are used in 327 and (327)$_0$ 
since strong variation of local interactions is not expected across these
systems.
An asterisk in Fig.~S1 corresponds to $(U,J_H)\sim(0.8,0.4)$, where units of $2t_{x}^{xz}=1$
in 214, given in Table SI,
are used throughout this work.

For the fixed values represented by the asterisk, the phase diagram in a magnetic
field perpendicular to the layers is shown in Fig.~\ref{pd}, where areas of 
finite SC are shown in red.
SC exists 214 and is increased when bilayer coupling is added in (327)$_0$.
The addition of staggered rotations in 327 destroys
SC, which is replaced by a paramagnet.  Further increasing the field eventually 
stabilizes SDW in 327, discussed below.  This overall tendency is independent of
the precise choice of $(U,J_H)$ 
because the evolution of SC to SDW originates from the reduction SC and 
enhancement of SDW in 327.

To understand why SC is enhanced in (327)$_0$,
consider the effect of the interlayer hopping in a two orbital model.
Interlayer hopping is orbital dependent and
largest
for $yz$ to $yz$ and $xz$ to $xz$ hopping,
while it is minimal between $xy$ orbitals. 
A two-orbital model of $xz$ (or $yz$) and $xy$ gives insight into the effects of
bilayer coupling.
We use the basis $\Psi_k^\dagger = (\psi_k^\dagger,\mathcal{T}\psi_k^T\mathcal{T}^{-1})$
where $\mathcal{T}$ represents time-reversal, and
$\psi_k^\dagger=(c_{k,1,\uparrow}^{xz\dagger},c_{k,1,\downarrow}^{xz\dagger},
c_{k,1,\uparrow}^{xy\dagger},c_{k,1,\downarrow}^{xy\dagger},
c_{k,2,\uparrow}^{xz\dagger},c_{k,2,\downarrow}^{xz\dagger},
c_{k,2,\uparrow}^{xy\dagger},c_{k,2,\downarrow}^{xy\dagger})$, 
where the top and bottom layers are represented by the subscript 1 and 2,
respectively.
The kinetic and SOC parts of the Hamiltonian are then given by,
\begin{equation}
\begin{aligned}
H_k = &~ \xi_k^+ \rho_3\eta_0\tau_0\sigma_0 + \xi_k^- \rho_3\eta_0 \tau_3 \sigma_0
+\lambda \rho_3\eta_0\tau_2\sigma_1 \\ &+ \frac{1}{2}t_\perp \rho_3\eta_1(\tau_0+\tau_3)\sigma_0,
\end{aligned}
\end{equation}
where $\eta$, $\tau$, $\sigma$, and $\rho$ are Pauli matrices representing the layer, orbital,
spin, and particle-hole bases, respectively.  
The orbital dispersions in $xz$ and $xy$
are represented by $\xi_k^{xz/xy}=\xi_k^+ \pm \xi_k^-$
and left in this general form here.  
Details of these dispersions used in the full three-orbital model are 
found in Appendix A.
The SOC is given by $\lambda$, 
and the interlayer hopping, $t_\perp$, exists only in $xz$.

The interorbital-singlet 
spin-triplet pairing, $\Delta_x^{xz/xy}$, with a $d$-vector in the $x$-direction is written as,
\begin{equation}
H_\text{SC}=\Delta_x^{xz/xy} \rho_1 \eta_0 \tau_2 \sigma_1.
\end{equation}
Transforming $H_\text{SC}$ to the band basis, i.e., the
basis in which $H_k$ is diagonal, it is clear  how intraband pairing forms, and how
the various terms affect the gap size;
the magnitude of the intraband gaps in the two bands is,
\begin{equation}
|\Delta_\text{band}^\pm|=\frac{2\lambda \Delta_x^{xz/xy}}{\sqrt{(\xi_k^-\pm \frac{1}{2}t_\perp)^2+4\lambda^2}}.
\end{equation}
This shows first that SOC is responsible for the formation of the intraband gap,
consistent with previous works when $t_\perp =0$ \cite{Clepkens2021PRR, Clepkens2021PRB}.
Second, 
it relates the interorbital SC pairing, $\Delta_x^{xz/xy}$, to
the pairing on the band, $\Delta_\text{band}$, where the pairing on 
the band is affected by the band dispersion $\sqrt{(\xi_k^- + \frac{1}{2}t_\perp)^2+4\lambda^2}$ 
or $\sqrt{(\xi_k^- - \frac{1}{2}t_\perp)^2+4\lambda^2}$.
The pairing on one of the bands is increased by $t_\perp$
if the degeneracy of the bands is increased.
Therefore, when the bilayer coupling brings two bands closer together
near the Fermi level, 
the resulting bands feature a more significant mixture of the two
orbitals with appropriate spin character via SOC, 
and thus are more ideal for interorbital SC to form.

\begin{figure}[t]
\includegraphics{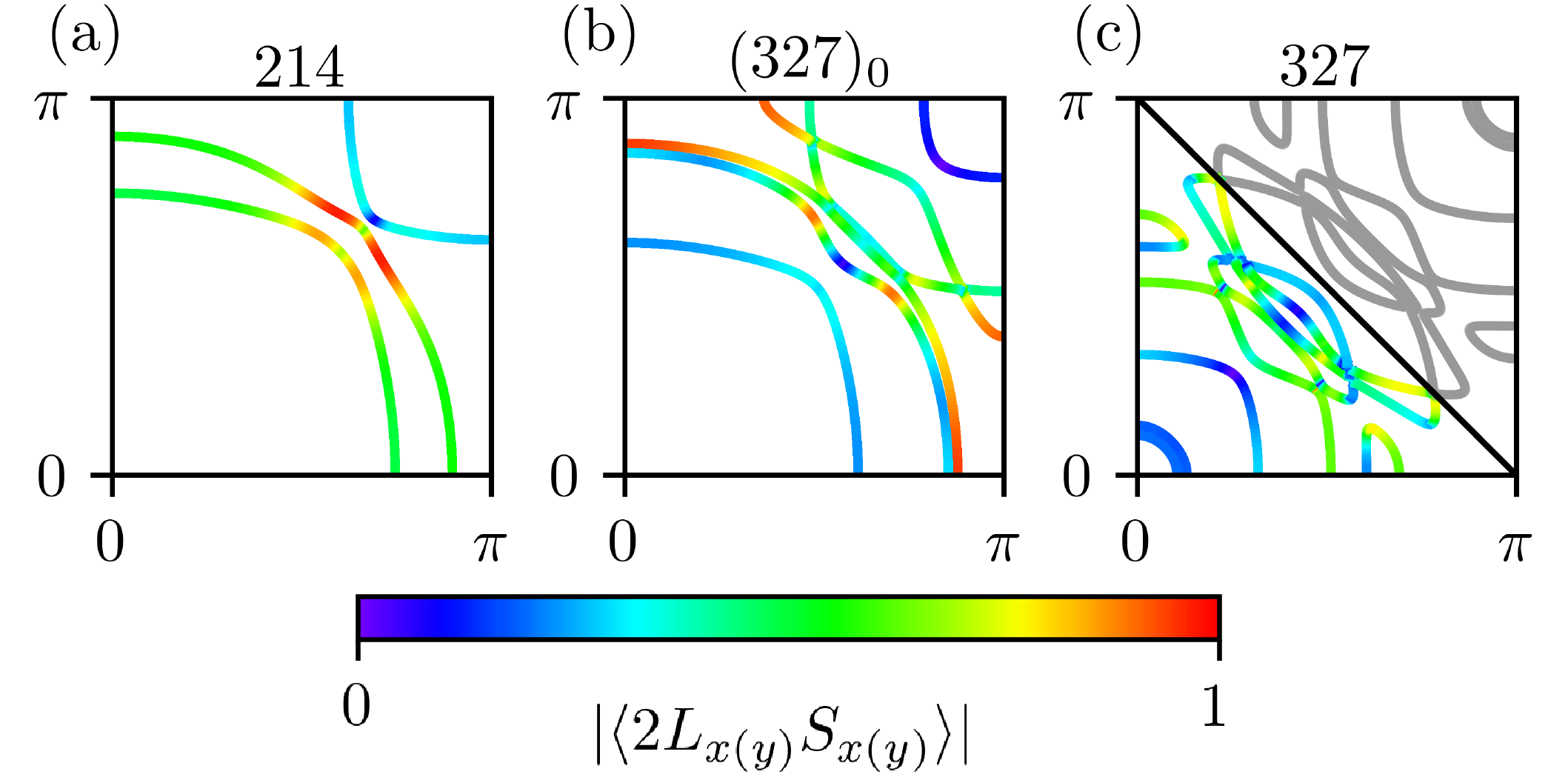}  
\caption{
Character of the mixing of orbitals and spins at the Fermi level at zero field, 
measured by $\langle2L_{x(y)}S_{x(y)}\rangle$, for
TB models of (a) 214, (b) the bilayer with no rotations,
$(327)_0$, and (c) 327. 
Larger mixing favors interorbital SC \cite{Puetter2012EPL}.
The areas where these states mix most are shown in red, where 
pairing will be strongest.
Introducing bilayer coupling that increase the band degeneracy leads to more 
areas where SC can form, as shown for (327)$_0$.
This is not the case when staggered rotations are included in 327
consistent with self-consistent MFA results shown in Fig.~\ref{pd}.
}\label{orbs2}
\end{figure}

The effect of the bilayer coupling on the orbital and spin mixing,
shown for 214 in Figs.~\ref{orbs2}(a) and (b),
is another 
way to understand the effect on interorbital SC pairing.
This displays the trend of orbital and spin mixing via SOC,
measured by $\langle 2L_xS_x\rangle$ or $\langle 2L_yS_y\rangle$,
favorable 
for $\Delta_{x}^{xz/xy}$ 
or $\Delta_y^{yz/xy}$, within each of the bands at the Fermi level.
Red areas are the most mixed, ideal for SC.
Note that not all bands exhibit the increase in such mixing due to
the bilayer, 
i.e., the bilayer can in principle decrease the band degeneracy,
which works against the SC.
For the 214 TB model though, we found that it works in favor of SC.

Next, we consider the effect of 
the staggered rotations present in 327.
While the bilayer coupling may bring band energies closer together and increase degeneracy, 
orbital hybridization is detrimental to the formation of interorbital-singlet spin-triplet SC 
\cite{Ramires2016PRB, Clepkens2021PRR}.
The hybridization from staggered rotations in 327 is included in 
Fig.~\ref{orbs2}(c), showing that areas of maximum mixing 
in the single layer and ideal bilayer have disappeared,
destroying SC.
This is consistent with the phase diagram of 327 in Fig.~\ref{pd}, 
obtained by self-consistent MFA calculations.

\section{Spin-density wave ordering}
In addition to interorbital SC, the Kanamori Hamiltonian also provides  
the intraorbital SDW instability. 
Specifically, with the Hubbard interaction,  SDW 
order appears 
when an appropriate nesting vector exists.
Previous studies on the magnetic susceptibility of 214
have found strong intraband nesting for 
$\textbf{q} \approx (\pm\frac{\pi}{2},\pm\pi)$ to
$(\pm\frac{2\pi}{3},\pm\pi)$ depending
on TB parameters \cite{Cobo2016PRB,Gingras2019PRL}.  
However, no SDW order is found in unstrained 214 at low temperatures, suggesting
that this nesting is not sufficient for SDW order.
Similar nesting is present in 327, with 
$\textbf{q} \approx (\pm\frac{\pi}{2},0)$ to $(\pm\frac{\pi}{3},0)$ due to the
unit cell doubling.  
In addition, there also exists the same vector rotated by $\frac{\pi}{2}$ (i.e., along the $y$-direction instead).
As in the case of 214, this nesting alone is not sufficient, 
as no SDW order forms at zero field.
It was shown that SDW order is suppressed by SOC in 327
by reducing nesting  \cite{Raghu2009PRB},
but the importance of SOC in this material has been recognized in 
explaining the dependence of the metamagnetic transition on the 
orientation of the field \cite{Raghu2009PRB}. 
Additionally, experiments show a strong dependence of the formation of 
SDW order on the orientation of the field \cite{Lester2015nmat}.

\begin{figure}[ht]
\includegraphics{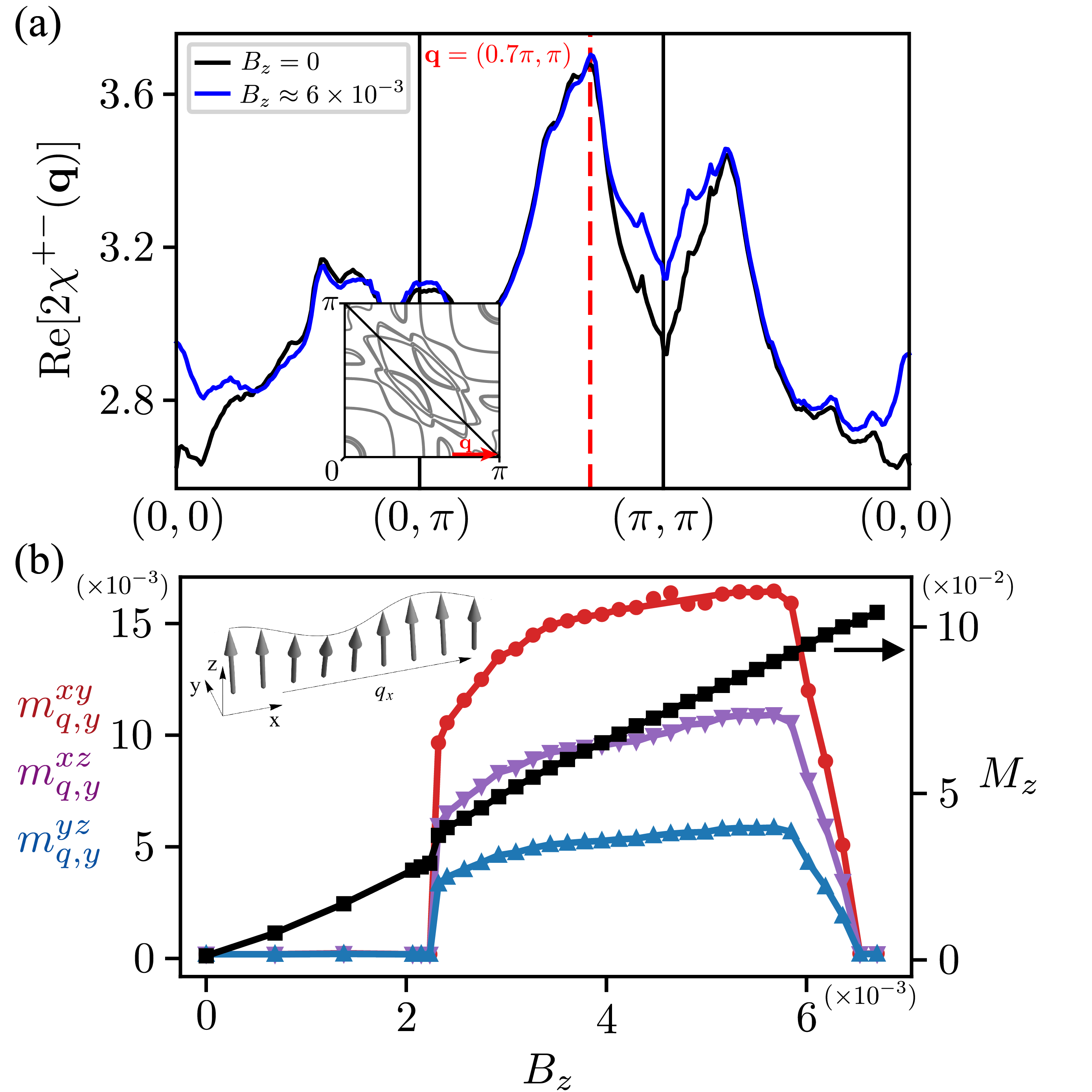}
\caption{
(a) Transverse susceptibility calculations plotted over the extended 
Brillouin zone with and without a field.  The peak at $\textbf{q}=(0.7\pi,\pi)$ increases
when the field is present and additional bands cross the Fermi level.
The Fermi surface
in the presence of a field (inset) shows the ordering wave-vector 
of the reduced BZ. 
(b) 
Magnitude of the SDW order parameters, $m_{q,y}^a$, where $a$ and $y$
represent the orbital and 
magnetic moment direction, respectively, 
and $M_z$, the magnitude of the $z$ magnetization.
The inset shows a pictorial representation of the spins in the SDW state, 
with an overall magnetization in the $z$-direction from the field, and an
oscillating $y$-component for $\textbf{q}$ in the $x$-direction.
}\label{FS}
\end{figure}

Although the nesting in 327 at zero field is not enough for SDW order to form, 
the presence of a van Hove singularity (vHS) was experimentally identified
in the bilayer in the presence of a field \cite{Tamai2008PRL}, 
and it was suggested that the metamagnetic and nematic transitions are
driven by the vHS
\cite{Kee2005PRB, Yamase2007jpsj, Lee2010PRB327, Puetter2010PRB, Efremov2019PRL,Mousatov2020pnas}.
We therefore consider the effect of nesting and vHSs together on 
the formation of SDW order in 327.
Analysis of the 327 TB model obtained via DFT reveals 
a peak in the DOS just below the Fermi level. 
Introducing a magnetic field splits the peak in two, 
leading to a significant change in the
magnetization once one peak reaches the Fermi level.
The Fermi surface under a field, shown in the inset of Fig.~\ref{FS}(a) shows 
the additional bands crossing the Fermi level near $(\pm\pi,0)$ and $(0,\pm\pi)$.

Considering possible ordering wave vectors, the intraorbital, intralayer transverse 
susceptibility is shown in Fig.~\ref{FS}(a). 
Importantly, the peak in the susceptibility increases when a field is present 
due to the additional bands crossing the Fermi level.
This ordering wave-vector corresponds to $\textbf{q}\approx(0.3\pi,0)$ in the reduced 
Brillouin zone, but slightly shifts as the field is changed.
Details of the susceptibility calculations are found in Appendix B.
To obtain numerical values, we perform self-consistent MFA calculations using 
Eq.~(\ref{Hint}) under a magnetic field.
The region of SDW order is shown in blue in Fig.~\ref{pd} where, 
for the choice of interaction and TB parameters, 
$B_{c1}\approx2.3\times10^{-3}$ and $B_{c2}\approx6.4\times10^{-3}$ with $\mu_B$ set to 1.
Setting $t_{xz}^x$ to the DFT obtained value,
$B_{c1}$ is of order 10 T. 
However, these values are sensitive to the location of
the vHSs in energy 
and the values of hopping parameters which can be further 
reduced by correlation effects beyond the MFA.
Thus the precise field values require further study beyond the scope of the
current work.

The SDW order parameters and magnetization are shown in 
Fig.~\ref{FS}(b), where 
the spin component is in the $y$-direction in each of the
three orbitals when the wave-vector is in the $x$-direction.
The primary contribution comes from the $xy$ orbital, and the 
secondary contribution from the $xz$ orbital.
The overall magnetization is in the $z$-direction with 
the metamagnetic transition occuring at $B_{c1}$.
A similar result is obtained when calculations are instead performed
using a wave-vector in the $y$-direction, the SDW order occurs 
with spin in the $x$-direction, with significant contributions from
the $xy$ then $yz$ orbitals. 
Within this model of the bilayer, no SC regions are found.

\section{Discussion and summary}
The absence of SC in 327 must be considered when explaining 
SC in 214 due to the similarity of the two materials.
The most significant differences, bilayer coupling
and staggered rotations, 
primarily affect interorbital SC 
and therefore provide a natural explanation for the
lack of SC in 327.
The presence of SDW order in 327
under a field
is another piece of the ruthenate puzzle.  The Kanamori Hamiltonian provides a
consistent framework for both SC and SDW order. The Hund's coupling  
plays an important role in stabilizing 
interorbital SC, while the Hubbard repulsion
and Hund's coupling together 
lead to SDW order.  
These interactions 
plus the increase in the DOS occurring in a field, experimentally observed as a 
metamagnetic transition, allow SDW ordering to form.
The interactions presented here are limited to on-site, 
but further neighbor interactions have been considered in interorbital SC
studies of 214 \cite{Romer2021PRB}. Their effect on  
SDW order remains to be studied. 

Some limitations of our theory arise from the finite size of the TB model
and MFA.
Experiments detect multiple metamagnetic transitions and two SDW phases with
different ordering wave-vectors all separated by less than 1T \cite{Tokiwa2016PRL}. 
Electron correlations beyond the MFA were shown to be important in the formation of the
heavy bands \cite{Allan2013iop},
important for
reproducing the narrow range over which the metamagnetic transitions and SDW phases occur.
The interorbital SC presented here requires $U'<J_H$ within the MFA.
It was shown by DMFT and other studies that such a requirement is not necessary beyond
the MFA \cite{Hoshino2015PRL,Hoshino2016PRB,Gingras2019PRL,Gingras2022PRL,Gingras2022PRB}.
To corroborate the earlier numerical studies, we calculated pairing correlators via
exact-diagonalization on a small cluster and found interorbital pairing tendency 
occurring when $J_H$ is about 20\% of $U$. 
This tendency is related to the changes of charge 
configurations via Hund's coupling, shown in Appendix D.

An area of recent development requiring consideration going forward is possible SDW order 
in strained 214 \cite{Grinenko2020}.  While no experimental observations have 
been reported about the ordering wave-vector or spin direction, this appears 
similar to 327 since SDW order occurs near a vHS. 
The shift of the vHS however, is induced by strain, so while
the $q_x$ and $q_y$ ordered states are degenerate in the bilayer, 
only one of these is chosen by the uniaxial strain.
Further work is required to 
fully differentiate the two cases. 
Other recent experiments have studied surface layers of 214 \cite{Marques2021},
and the trilayer, Sr$_4$Ru$_3$O$_{10}$ \cite{Gebreyesus2021}, both of which
feature staggered rotations. 
No SC is found in either, consistent
with the current theory.
However, future works are needed to estimate surface effects 
including broken inversion symmetry.

In summary, we showed that interorbital SC provides a natural
explanation for both the presence of SC in 214 and
lack of SC in 327.
In 327,  SC is destroyed by orbital hybridization
introduced by staggered rotations, 
not present in 214.
This same model finds intraorbital SDW ordering in 327 when the vHS 
reaches the Fermi level via a magnetic field.
We believe that the lack of SC in 327 is an important 
piece of the puzzle to understand  SC in 
214, and therefore interorbital SC should remain 
among the promising candidates. 
Additionally, we predict that ideal (327)$_0$ 
can feature increased orbital degeneracy
by tuning the bilayer coupling strength, and therefore can also exhibit
interorbital SC with a possibly higher transition temperature.

\begin{acknowledgments} 
We would like to thank P.~Stavropoulos for useful discussion. 
This work was supported by the Natural Sciences and Engineering Research Council of Canada Discovery Grant 06089-2016, and the Center for Quantum Materials at the University of Toronto.
Computations were performed on the Niagara supercomputer
at the SciNet HPC Consortium. SciNet is funded by: the
Canada Foundation for Innovation under the auspices of
Compute Canada; the Government of Ontario; Ontario
Research Fund - Research Excellence; and the University
of Toronto.
\end{acknowledgments} 

\bibliography{SC-SDW}

\appendix

\section{Tight-binding model}

To obtain tight-binding parameters, we perform density functional theory
calculations with the Vienna \textit{ab initio} simulation package (VASP)
\cite{VASP}, using the projector augmented-wave potential \cite{paw} 
and the Perdew-Burke-Erzenhof exchange-correlation functional \cite{pbe}, 
with an energy cutoff of 400 eV.  The relevant TB parameters are obtained
using WANNIER90 \cite{Wannier}, and are listed in Tables \ref{tb1} and
\ref{tb2}.

The TB model is given by,
\begin{widetext}
\begin{equation}
\begin{aligned}
H_k =&~ \sum_{k,a,l} \xi_k^a c_{k,l,\sigma}^{a\dagger} c_{k,l,\sigma}^a
+\sum_{k,l} t_k c_{k,l,\sigma}^{xz \dagger} c_{k,l,\sigma}^{yz} + \text{h.c.} 
+\sum_{k} t_\perp^{1D}(c_{k,t,\sigma}^{yz\dagger}c_{k,b,\sigma}^{yz}+
c_{k,t,\sigma}^{xz\dagger}c_{k,b,\sigma}^{xz} ) \\
&+ t_\perp^{xy} c_{k,t,\sigma}^{xy\dagger}c_{k,b,\sigma}^{xy} - 
2t_{\perp,x}^{xz}(c_{k,t,\sigma}^{yz\dagger}c_{k,b,\sigma}^{yz}\cos k_y 
+c_{k,t,\sigma}^{xz\dagger}c_{k,b,\sigma}^{xz}\cos k_x)
- 2t_{\perp,x}^{xz/xy}(\sin k_y [c_{k,b,\sigma}^{xy\dagger}c_{k,t,\sigma}^{xz} \\ 
&-c_{k,t,\sigma}^{xy\dagger}c_{k,b,\sigma}^{xz}] + \sin_x[c_{k,b,\sigma}^{xy\dagger}c_{k,t,\sigma}^{yz} 
-c_{k,t,\sigma}^{xy\dagger}c_{k,b,\sigma}^{yz}])
+ \text{h.c.} \\ 
&+\sum_k 2t^\text{stag}(\cos k_x + \cos k_y) (c_{k,t,\sigma}^{yz\dagger}c_{k+Q,t,\sigma}^{xz} 
-c_{k,t,\sigma}^{xz\dagger}c_{k+Q,t,\sigma}^{yz}
- c_{k,b,\sigma}^{yz\dagger}c_{k+Q,b,\sigma}^{xz} 
+c_{k,b,\sigma}^{xz\dagger}c_{k+Q,b,\sigma}^{yz}
) + \text{h.c.} \\
&+\sum_k t_\perp^\text{stag} (c_{k,t,\sigma}^{yz\dagger}c_{k+Q,b,\sigma}^{xz}
-c_{k,t,\sigma}^{xz\dagger}c_{k+Q,b,\sigma}^{yz}) + \text{h.c.}.
\end{aligned}
\end{equation}
\end{widetext}
Where $a\in(yz,xz,xy)$ represents the orbitals, and $l\in(t,b)$ is the 
layer index.
The intraorbital dispersions are,
\begin{equation}
\begin{aligned}
\xi_k^{xz(yz)} =&~ -\mu_{1D} -2t_x^{xz}\cos k_{x(y)} - 2t_y^{xz}\cos k_{y(x)} \\
&-4t_{x,y}^{xz}\cos k_x  \cos k_y
-2t_{2x}^{xz}\cos 2k_{x(y)} \\ 
&- 2t_{3x}^{xz} \cos 3k_{x(y)} 
- 4t_{2x,y}^{xz}\cos 2k_{x(y)} \cos k_{y(x)},
\end{aligned}
\end{equation}
\begin{equation}
\begin{aligned}
\xi_k^{xy} =&~ 
-\mu_{xy} - \sum_{n} [2t_{nx}^{xy}(\cos nk_{x} + \cos nk_{y}) \\
&+ 4t_{nx,ny}^{xy}\cos nk_x \cos nk_y] \\
&- \sum_{m\ne n} 4t_{mx,ny}^{xy} (\cos mk_x \cos nk_y 
+\cos nk_x \cos mk_y)
\end{aligned}
\end{equation}
where $m$ and $n$ are the integers listed in Table \ref{tb2} describing
the intraorbital hopping within the $xy$ orbitals between sites separated 
by $m\hat{x}+n\hat{y}$ as well as $n\hat{x}+m\hat{y}$ by symmetry.  
The dispersion of the $xy$ orbital includes further neighbor hopping 
terms to model the flatness of the bands near $(\pm\pi,0)$ and $(0,\pm\pi)$ of
the Brillouin zone in the real bilayer material.
The orbital mixing is given by,
\begin{equation}
t_k = -4t_{1D}\sin k_x \sin k_y.
\end{equation}
The interlayer hoppings, $t_\perp^{1D}$ and $t_\perp^{xy}$, represent 
hopping between layers within orbitals directly above or below each other,
while $t_{\perp,x}^{xz}$ is hopping in the $xz(yz)$ orbital between
layers and one lattice site over in the $x(y)$-direction, and $t_{\perp,y}^{xz/xy}$ 
hopping between layers, one lattice site over in the $y(x)$-direction, between $xy$ and $xz(yz)$ orbitals.
The staggered hopping, $t^\text{stag}$, allowed by the staggered rotation, 
represents the hopping between NN $1D$
orbitals within a layer and changes sign between layers due to the opposite 
sense of the octahedra rotations between the layers, while $t_\perp^\text{stag}$
occurs between $1D$ orbitals directly above or below each other.
Finally, the SOC Hamiltonian is,
\begin{equation}\label{soc}
H_{\text{SOC}}=i\lambda \sum_i \sum_{abl} \epsilon_{abl} c_{i\sigma}^{a\dagger}c_{i\sigma'}^b \hat{\sigma}_{\sigma\sigma'}^l,
\end{equation}
where the strength of the SOC used is set to $\lambda=62.5$ meV throughout this work.

\begin{table}[htp]
\begin{center}
\begin{tabularx}{\columnwidth}{|>{\centering\arraybackslash}X >{\centering\arraybackslash}X >{\centering\arraybackslash}X >{\centering\arraybackslash}X >{\centering\arraybackslash}X|}
    \hline
    $t_{x}^{xz}$ & $t_y^{xz}$ & $t_{x,y}^{xz}$ & $t_{2x}^{xz}$ & $t_{2x,y}^{xz}$ \\
    \hline
    290.8 & 44.8 & $-16.8$ & $-47.0$ & $-13.5$ \\ 
    \hline\hline
    $t_{3x}^{xz}$ & $t_{1D}$ & $\mu_{1D}$ & $t_\perp^{1D}$ & $t_{\perp,x}^{xz}$ \\ 
    \hline
    $-3.2$ & 10.1 & 295.9 & (290.8) & (120.3) \\
    \hline\hline
    $t_{x}^{xy}$ & $t_{x,y}^{xy}$ & $t_{2x}^{xy}$ & $t_{2x,y}^{xy}$ & $t_{2x,2y}^{xy}$ \\
    \hline
    369.6 & 120.3 & $-4.2$ & 20.3 & 14.1 \\ 
    \hline\hline
    $t_{3x}^{xy}$ & $t_{3x,y}^{xy}$ & $t_{3x,2y}^{xy}$ & $t_{4x}^{xy}$ & $t_{3x,3y}^{xy}$ \\
    \hline
    4.8 & 4.4 & 5.5 & 1.9 & 4.0 \\ 
    \hline\hline
    $t_{4x,y}^{xy}$ & $t_{4x,2y}^{xy}$ & $t_{4x,3y}^{xy}$ & $t_{5x}^{xy}$ & $t_{5x,y}^{xy}$ \\
    \hline
    2.6 & 3.1 & 2.9 & x & x \\ 
    \hline\hline
    $\mu_{xy}$ & $t_{\perp}^{xy}$ & $t_{\perp,y}^{xz,xy}$ & $t^{\text{stag}}$&$t_{\perp}^{\text{stag}}$ \\
    \hline
    426.4 & (44.8) & (10.1) & 0 & 0 \\ 
    \hline
\end{tabularx}
\end{center}
\caption{Tight-binding parameters obtained for the single bilayer, 214,
plus the additional bilayer coupling terms used to model the ideal bilayer, 
$(327)_0$, shown in parentheses.
Note that the single layer and the ideal bilayer do not feature staggered 
rotations of the octahedra, so the staggered hopping terms are all 0 for both.
Values that were not able to be obtained from DFT results are marked with an `x'. 
While these values are important when considering the vHS in the real 
bilayer, they are not expected to have any significant effect in the single layer since 
flat bands near the Fermi level are not present.
All values are shown in units of meV.
Values listed in the main text are all given in units of 
$2t_{x}^{xz} = 1$.
}\label{tb1}
\end{table}

In calculations which feature a magnetic field in the $z$-direction, 
the additional term included in the
Hamiltonian is,
\begin{equation}
H_B = -B_z \sum_i (2S_i^z + L_i^z).
\end{equation}
The total $z$-magnetization calculated in the presence of the field is given by,
\begin{equation}
M_z = \sum_i \langle 2S_i^z + L_i^z\rangle
\end{equation}

\begin{table}[H]
\begin{center}
\begin{tabularx}{\columnwidth}{|>{\centering\arraybackslash}X >{\centering\arraybackslash}X >{\centering\arraybackslash}X >{\centering\arraybackslash}X >{\centering\arraybackslash}X|}
    \hline
    $t_{x}^{xz}$ & $t_y^{xz}$ & $t_{x,y}^{xz}$ & $t_{2x}^{xz}$ & $t_{2x,y}^{xz}$ \\
    \hline
    270.2 & 12.1 & $-15.1$ & $-29.8$ & $-10.1$ \\ 
    \hline\hline
    $t_{3x}^{xz}$ & $t_{1D}$ & $\mu_{1D}$ & $t_\perp^{1D}$ & $t_{\perp,x}^{xz}$ \\ 
    \hline
    $-3.5$ & 12.6 & 282.9 & 281.6 & 90.6\\
    \hline\hline
    $t_{x}^{xy}$ & $t_{x,y}^{xy}$ & $t_{2x}^{xy}$ & $t_{2x,y}^{xy}$ & $t_{2x,2y}^{xy}$ \\
    \hline
    294.3 & 137.2 & 32.5 & 13.6 & 22.1 \\ 
    \hline\hline
    $t_{3x}^{xy}$ & $t_{3x,y}^{xy}$ & $t_{3x,2y}^{xy}$ & $t_{4x}^{xy}$ & $t_{3x,3y}^{xy}$ \\
    \hline
    $-12.5$ & 10.3 & 1.34 & 8.99 & 2.2 \\ 
    \hline\hline
    $t_{4x,y}^{xy}$ & $t_{4x,2y}^{xy}$ & $t_{4x,3y}^{xy}$ & $t_{5x}^{xy}$ & $t_{5x,y}^{xy}$ \\
    \hline
    $-1.9$ & 1.1 & 1.3 & $-3.76$ & 0.7 \\ 
    \hline\hline
    $\mu_{xy}$ & $t_{\perp}^{xy}$ & $t_{\perp,y}^{xz,xy}$ & $t^{\text{stag}}$&$t_{\perp}^{\text{stag}}$ \\
    \hline
    400.7 & 18.6 & 13.6 & 84.5 & 64.3 \\ 
    \hline

\end{tabularx}
\end{center}
\caption{Tight-binding parameters obtained for the real bilayer, 327.
All values are shown in units of meV.
Note that the chemical potential listed here is slightly below a filling of 2/3 to 
ensure that the vHS is close enough to the Fermi level to produce the metamagnetic 
transition at a field value near the experimentally observed 7.9 T.
The chemical potential used corresponds to a filling fraction of approximately 0.657.
}\label{tb2}
\end{table}

\section{Susceptibility}

Defining the components of the spin operator for orbital $\alpha = (yz,xz,xy)$ and layer $l = (t, b)$ as,
\begin{equation}
    S^{a}_{\alpha l}(\textbf{q}) = \sum_{\textbf{k},
    \sigma\sigma'}c^{\dagger}_{\textbf{k}\alpha l\sigma}\frac{\sigma^{a}_{\sigma\sigma'}}{2}c_{\textbf{k}+\textbf{q}\alpha l\sigma'},
\end{equation}
the total bare spin susceptibility is calculated by summing over the orbital and layer indices in the Fourier transform of the imaginary time spin-spin correlation function
\begin{equation}
\begin{aligned}
    &\chi^{ab}(\textbf{q}, i\omega_{n}) = \sum_{\alpha,\beta,l,m}[\chi^{ab}]^{\alpha l}_{\beta m}(\textbf{q},i\omega_{n}),
    \\[6pt]&[\chi^{ab}]^{\alpha l}_{\beta m}(\textbf{q}, i\omega_{n}) = \frac{1}{N}\int_{0}^{\beta}d\tau e^{i\omega_{n}\tau}\langle T_{\tau}S^{a}_{\alpha l}(\textbf{q}, \tau)S^{b}_{\beta m}(-\textbf{q}, 0)\rangle
\end{aligned}
\end{equation}
where $\beta = \frac{1}{k_{B}T}$, $\tau$ is imaginary time and $T_{\tau}$ is the imaginary time-ordering operator. To evaluate this we use the transformation to the band basis given by, $c_{\textbf{k}n} = \sum_{s}u_{ns}(\textbf{k})d_{\textbf{k}s}$, where $n = (\alpha,l,\sigma)$, $s$ indexes the bands, and $u_{ns}(\textbf{k})$ are the coefficients of the unitary transformation. Evaluating the correlation function in terms of the Green's functions in the band basis and subsequently the Matsubara sum via contour integration, with the analytic continuation $i\omega_{n}\rightarrow w+i\eta$, the components of the static susceptibility are given by
\begin{widetext}
\begin{equation}
\begin{aligned}
    [\chi^{ab}]^{\alpha l}_{\beta m}(\textbf{q}, 0) = -\frac{1}{4N}\sum_{\textbf{k},s,s',\sigma_{1},\sigma_{2},\sigma_{3},\sigma_{4}}&u^{*}_{\alpha l\sigma_{1},s}(\textbf{k})\sigma^{a}_{\sigma_{1}\sigma_{2}}u_{\alpha l\sigma_{2},s'}(\textbf{k}+\textbf{q})u^{*}_{\beta m\sigma_{3},s'}(\textbf{k}+\textbf{q})\sigma^{b}_{\sigma_{3}\sigma_{4}}u_{\beta m\sigma_{4},s}(\textbf{k})\\\times~ &\frac{f(\xi_{\textbf{k}}^{s})-f(\xi_{\textbf{k}+\textbf{q}}^{s'})}{ \xi_{\textbf{k}}^{s}-\xi_{\textbf{k}+\textbf{q}}^{s'}+ i\eta}.
\end{aligned}
\end{equation}
\end{widetext}
Below we show the intra-layer contribution to the transverse susceptibility from each orbital, $\sum_{l}[\chi^{+-}]^{\alpha l}_{\alpha l}(\textbf{q})$, the intra-layer contribution summed over the orbital index, $\sum_{\alpha,\beta,l}[\chi^{+-}]^{\alpha l}_{\beta l}(\textbf{q})$, and the inter-layer contribution summed over the orbital index, $\sum_{\alpha,\beta,l\neq m}[\chi^{+-}]^{\alpha l}_{\beta m}(\textbf{q})$.


\begin{figure}[htp]
\includegraphics[width=\columnwidth]{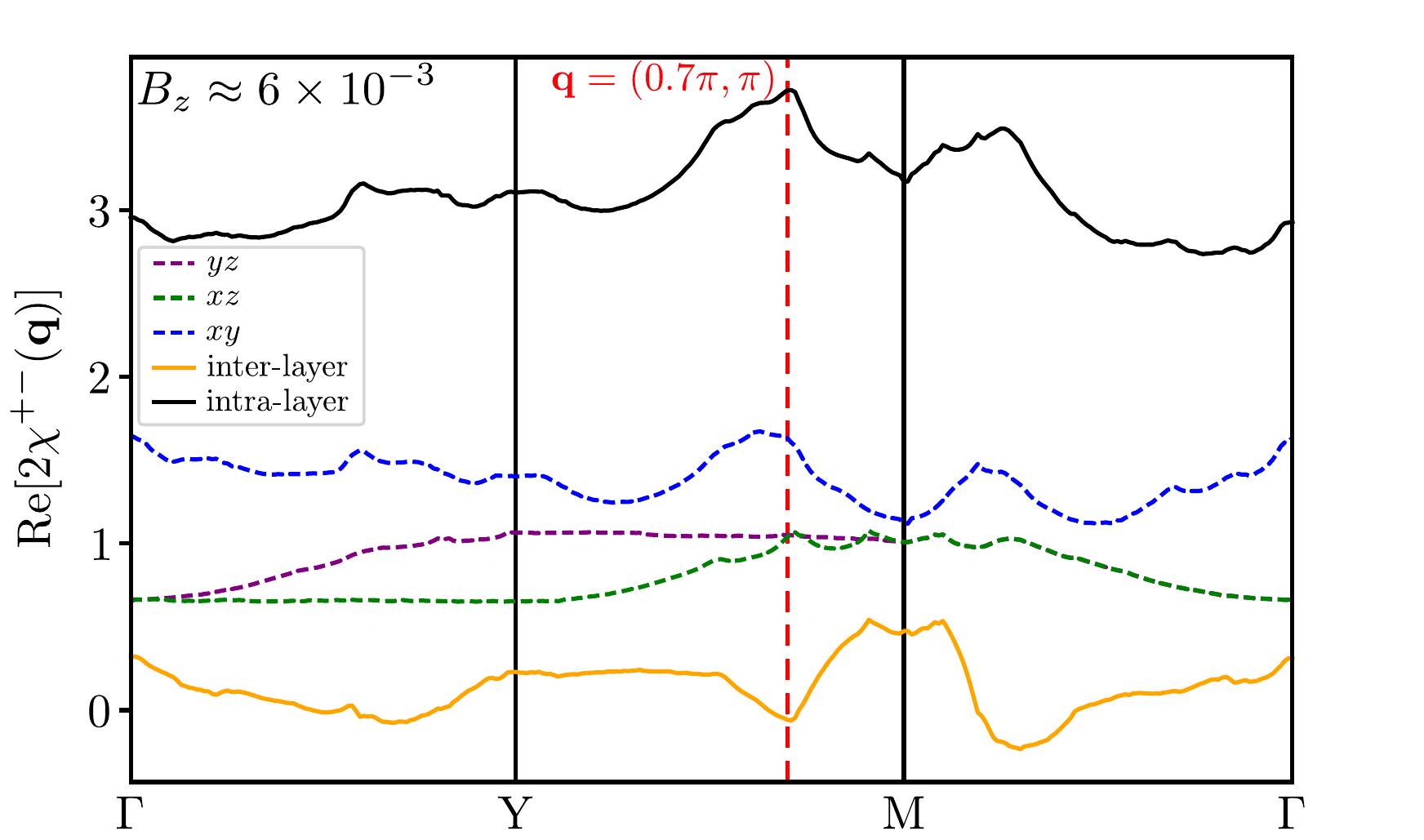}
\caption{
Transverse susceptibility of the 327 model in the presence of a finite field.
The intraorbital, intralayer susceptibilities are individually plotted, 
which provide the primary contributions to the ordering. The
total intralayer susceptibility including intraorbital and interorbital 
contributions is also shown.
Additionally, the total \textit{inter}layer susceptibility is plotted, which, when
included may modify the ordering wave vector 
due to the differing locations of its peaks.
}\label{sus}
\end{figure}

\section{Full Phase Diagram}

\begin{figure}[htp]
\includegraphics{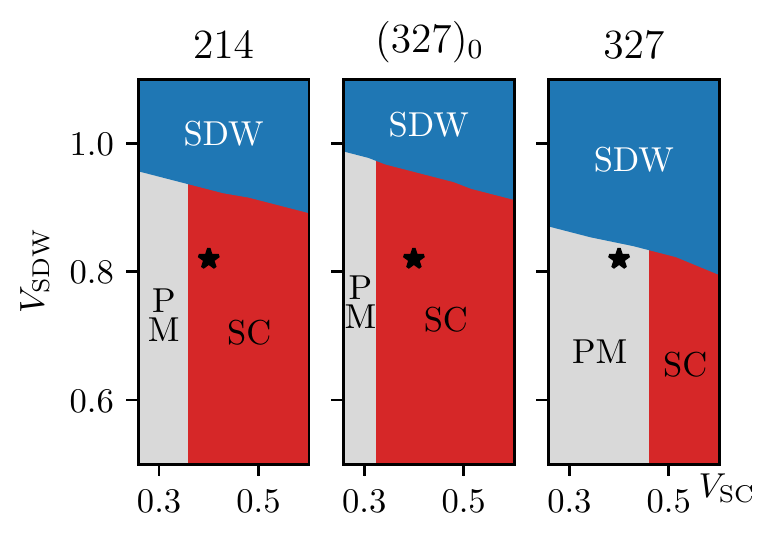}
\caption{
Full phase diagram of Eq.~(1) from the main text for each of the TB 
sets in Tables.~\ref{tb1} and \ref{tb2}, where $V_\text{SDW} = U$ and 
$V_\text{SC}=3J_H-U$.  All calculations are performed
without a field.  In the presence of a field, the SDW region of the 327
model expands due to the vHSs crossing the Fermi level.  The asterisks
correspond to the values used in the calculations presented in the main
text.
}\label{2dpd}
\end{figure}

To show the behaviour of all three models considered for various sets of 
interaction parameters, the full phase diagrams are shown in Fig.~\ref{2dpd}.
The superconducting region expands when the bilayer coupling is added to the single layer
model, however, the superconducting region shrinks and the spin-density wave region
is expanded when the staggered rotations are also added.
The asterisks correspond to the values used in the main text, where superconductivity 
exists in 214 and $(327)_0$, while 327 remains a paramagnet at zero field and only
exhibits SDW order when a field is introduced.

\section{Exact Diagonalization Correlators}

\begin{figure}[htp]
\includegraphics[width=\columnwidth]{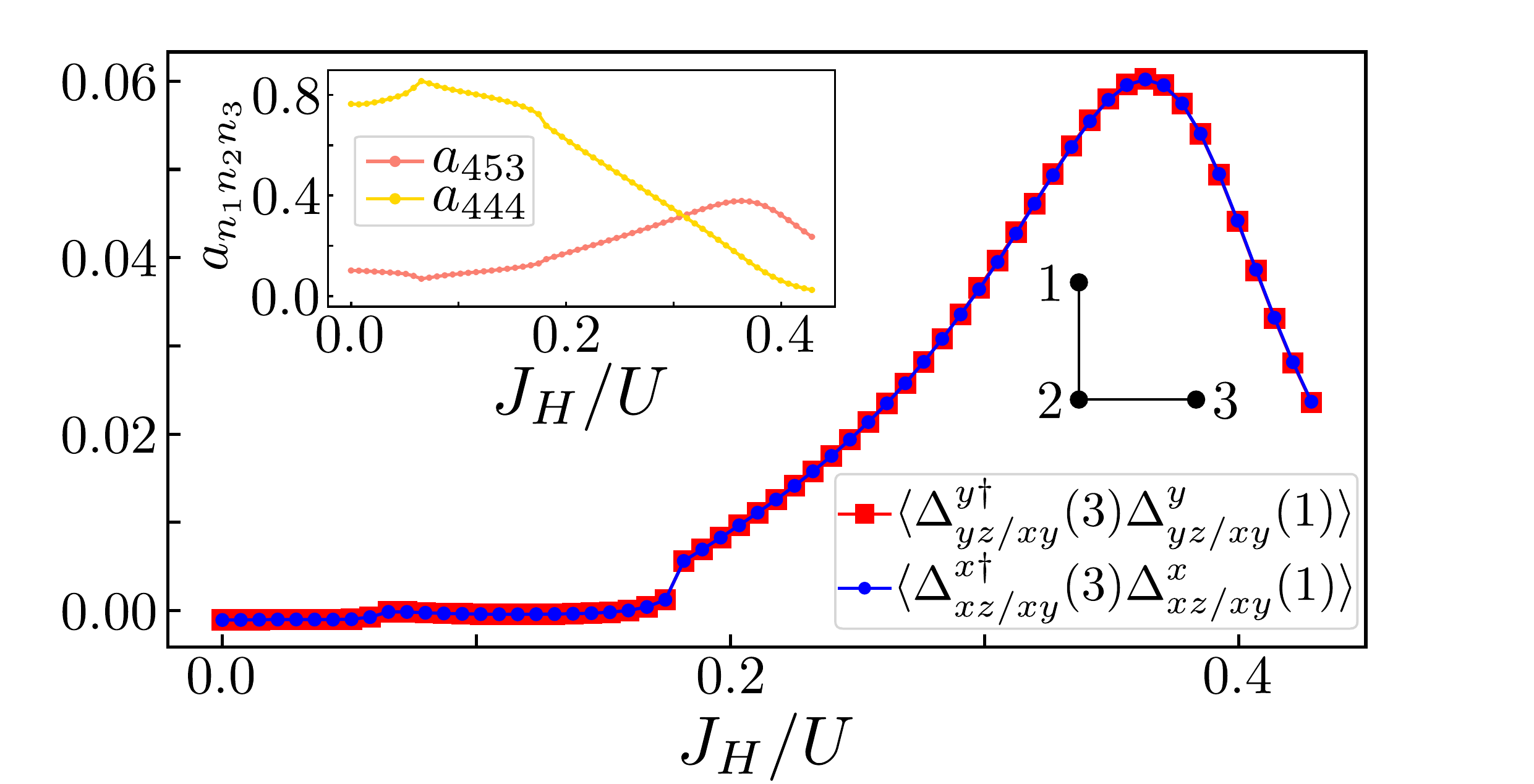}
\caption{
Pairing correlators from three-site exact diagonalization calculations as a function
of $J_H/U$, calculated for 
the 214 tight-binding parameters shown in Table \ref{tb1}, and the Kanamori Hamiltonian using $U=3.5$.
The correlators are finite before the MFA required minimum ratio of $J_H/U>1/3$, 
suggesting that electron fluctuations can reduce the required Hund's coupling 
necessary for SC to form.
The electron configurations are shown in the upper left inset, where $a_{n_1n_2n_3}$ refers to the 
weight of $d_{1}^{n_1}d_2^{n_2}d_3^{n_3}$ configuration in the wavefunction,
where $n_i$ represents the number of electrons at site $i$.
The pairing correlators become finite at the interaction ratio corresponding to the sudden drop
of the weight of the $d_1^4d_2^4d_3^4$ configuration
and increase in the $d_1^4d_2^5d_3^3$ configuration, boosted by Hund's coupling.
The $a_{444}$ and $a_{453}$ weights are shown, but other configurations also 
contribute to the wavefunction.
The geometry of the three-site configuration is shown under the curve.
}\label{cor}
\end{figure}

To extend the consideration of SC beyond the MFA, in Fig.~\ref{cor}, 
we show an exact diagonalization calculation for a three-site cluster
as shown by the dots below the curve in Fig.~\ref{cor}, containing three $t_{2g}$ orbitals per site and 12 total electrons.
Including spin the dimension of the Hilbert space is $\binom{18}{12} = 18564$.
The Hamiltonian includes the same Kanamori interaction terms considered in the main text
and the ground state is obtained with the Lanczos algorithm with a convergence tolerance of $10^{-15}$.
The pairing correlation function, $\langle \Delta_{a/b}^{l\dagger}(i)\Delta_{a/b}^{l}(j)\rangle$, 
for the interorbital-singlet spin-triplet pairings considered in the main text is evaluated between
the two next-nearest-neighbour sites separated along the diagonal ($i=3, j=1)$ as a function of 
$J_{H}/U$ with the hopping parameters for 214 shown in Table \ref{tb1} and $U=3.5$.
The pairing correlations for the $\Delta^{x}_{xz/xy}$ and $\Delta^{y}_{yz/xy}$ operators
start to increase at $J_{H}/U \approx 0.18$, and are matched by an increase in charge 
fluctuations away from the $d_1^{4}d_2^{4}d_3^{4}$ configuration to 
configurations such as $d_1^{4}d_2^{5}d_3^{3}$,
where $d_i^n$ refers to $n$ electrons at site $i$.
The weight of the $d_1^4d_2^4d_3^4$ and $d_1^4d_2^5d_3^3$ configurations in the ground state wavefunction
are shown in the upper left inset of Fig.~\ref{cor}, denoted $a_{444}$ and $a_{453}$, respectively.
Other configurations also contribute to the ground state wavefunction, but are omitted here.
The inclusion of electron fluctuations as allowed by these calculations shows
that the strict requirement of $J_H/U>1/3$ is only necessary in the MFA where
the effects of charge configurations such as $d_1^4d_2^5d_3^3$ combined 
with Hund's coupling 
cannot be taken into account.


\end{document}